\documentclass[aip,reprint]{revtex4-1}

\usepackage{graphicx}
\usepackage{dcolumn}
\usepackage{bm}
\usepackage{gensymb}
\usepackage[utf8]{inputenc}
\usepackage[T1]{fontenc}
\usepackage{mathptmx}
\usepackage{etoolbox}
\usepackage[english]{babel} 
\makeatletter
\def\@email#1#2{%
 \endgroup
 \patchcmd{\titleblock@produce}
  {\frontmatter@RRAPformat}
  {\frontmatter@RRAPformat{\produce@RRAP{*#1\href{mailto:#2}{#2}}}\frontmatter@RRAPformat}
  {}{}
}%
\makeatother

\draft 

\begin{document}

\title{Phononic Combs in Lithium Niobate Acoustic Resonators}

\author{I. Anderson}

\email{ianderson@utexas.edu}

\author{J. Kramer} \author{T.H. Hsu} \author{Y. Wang} \author{V. Chulukhadze} \author{R. Lu}

\affiliation{ 
 Department of Electrical and Computer Engineering, University of Texas at Austin, Austin, TX 78758, USA
}

\date{\today}
             
\begin{abstract}
Frequency combs consist of a spectrum of evenly spaced spectral lines. Optical frequency combs enable technologies ranging from timing, LiDAR, and ultra-stable signal sources. Microwave frequency combs are analogous to optical frequency combs, but often leverage electronic nonlinearity for comb generation. Generating microwave frequency combs using piezoelectric mechanical resonators would enable this behavior in a more compact form factor, thanks to the shorter acoustic wavelengths. In this work, we demonstrate a microwave frequency comb leveraging thermal nonlinearity in high quality factor ($Q$) overmoded acoustic resonators in thin film lithium niobate. By providing input power at 257 MHz, which is the sum frequency of two acoustic modes at 86 MHz and 171 MHz, we generate parametric down conversion and comb generation. We explore the nonlinear mixing regimes and the associated conditions for comb generation. Comb spacing is observed to vary significantly with drive frequency and power, and its general behavior is found to rely heavily on initial conditions. This demonstration showcases the potential for further improvement in compact and efficient microwave frequency combs, leveraging nonlinear acoustic resonators. 
\end{abstract}

\maketitle 

\begin{figure*}
\includegraphics{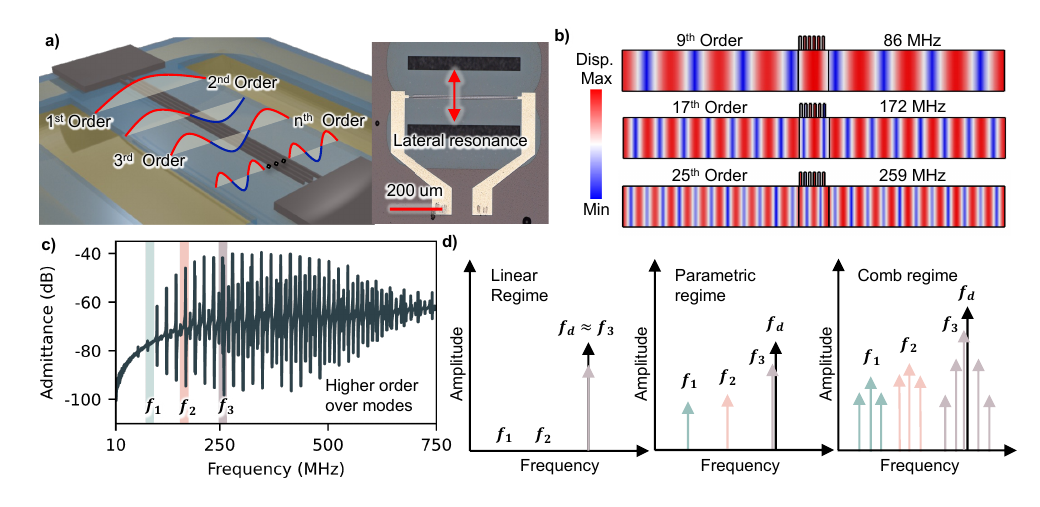}
\caption{\label{fig:wide1}Device overview detailing design and operation. (a) 3D model and optical image of device highlighting the resonance conditions and directions. (b) COMSOL FEA simulations highlighting the high Q modes that are used for our devices. (c) Measured admittance response of device, with the specific modes again highlighted. (d) schematic of different regimes of driving for comb generation, consisting of linear, parametric down-conversion, and then comb regime.}
\end{figure*}

Frequency combs rely on nonlinear effects to produce evenly spaced spectral lines around a central frequency.  When these lines are coherently added in the time domain, they form a train of ultrashort pulses compared to the original input signal. Both frequency and time domain waveforms provide benefits for creating and controlling signals. For example, optical frequency combs have been used for ultrashort pulse generation via mode-locked lasers or through Kerr-based microresonators \cite{delhaye_optical_2007,fortier_20_2019}. These combs have enabled transformative applications such as dual-comb spectroscopy, LiDAR, and the development of highly accurate, ultra-stable signal sources for high precision measurements. \cite{benedick_compact_2009,schliesser_frequency-comb_2005,long_electro-optic_2021,shimizu_optical_2020}.

The attention garnered by optical frequency combs has led researchers to pursue such phenomena in the mechanical domain. Mechanical nonlinearities allow for an analogous system capable of generating frequency combs \cite{kovacic_duffing_2011,van_der_avoort_parametric_2011,shao_nonlinearity_2008}. Termed phononic combs, these devices were first shown in electrically actuated extensional mode resonators in aluminum nitride on silicon using laser doppler vibrometer (LDV)\cite{ganesan_phononic_2017}. Different from typical Kerr combs, these phononic frequency combs typically require parametric down-conversion, and the beat note frequency, or comb spacing, arises from the offset between the sum of frequencies of the parametric tones and the frequency of the driven mode in a cascaded three wave mixing process. These microelectromechanical systems (MEMS) combs have since been realized in a number of material systems and resonator structures, providing an opportunity for microwave frequency combs generation in compact structures \cite{maksymov_acoustic_2022}. More traditional single mode Kerr-combs have also been demonstrated in systems such as quartz\cite{goryachev_generation_2020,kubena_phononic_2020}, silicon\cite{czaplewski_bifurcation_2018,wang_frequency_2022}, and similar platforms such as optomechanical systems\cite{zhang_optomechanical_2021,rahmanian_elastomeric_2025}.  These microwave frequency combs offer a new route for applications such as energy harvesting, range finding, and fundamental mode excitation\cite{zhang_mems_2023,hussein_passive_2024,antonio_frequency_2012}. 
In this work, we show the first demonstration of a phononic comb in lithium niobate (LN) using lamb wave resonators (LWR) above 100 MHz. These devices are driven with high power ($\sim$15 dBm) to first generate parametric down-conversion from one mode to two lower frequency modes. Then, at even higher power, they are then driven into the comb regime where each tone contains equidistant lines. We believe this is also the first comb generated using thermal nonlinearities rather than mechanical, which, while advantageous in terms of generating combs for micro-scale devices that typically don't have access to mechanical nonlinearities, causes said combs to be highly sensitive to initial conditions.

While the origin of nonlinearity differs from that of other phononic combs, the mathematical description and methods for generation still follow a very similar route to those previously reported. The dynamics of said comb are described by the equation below, where each of the 3 mechanical modes is coupled by various nonlinear terms, following the formalism shown in Ref.~\onlinecite{ganesan_excitation_2018}.
\begin{eqnarray}
\nonumber \ddot{Q}_i&=&-\omega_i^2Q_i-2\zeta_i\omega_i\dot{Q_i}
+\sum_{\tau_1}^{3}\sum_{\tau_2}^{3}\alpha_{\tau_1\tau_2}Q_{\tau_1}Q_{\tau_2} \\
&&+\sum_{\tau_1}^{3}\sum_{\tau_2}^{3}\sum_{\tau_3}^{3}\beta_{\tau_1\tau_2\tau_3}Q_{\tau_1}Q_{\tau_2}Q_{\tau_3}+P\cos(\omega_d t)
\end{eqnarray}
where $i$= 1, 2, or 3, $Q_i$ represents our mode profile, $\omega_i/\omega_d$ our mode and drive frequencies, $\zeta_i$ our damping coefficient, $\alpha$ and $\beta$ our second and third order nonlinearities, and $P$ our drive power. Combined, these equations represent three mechanical oscillators with varying nonlinear coupling terms of several orders that convert energy from one mode to another. The following equations are truncated to third order nonlinearity, but can certainly include higher orders \cite{ganesan_towards_2017}. The exact values of the coefficients in the above equation, able to be gathered from the experimental data, will be discussed in much more depth in future works more closely focused on theoretical modeling.

The generation of combs follows the process described below. Acoustic resonators are first designed in X-Cut LN targeting Lateral Overtone Bulk Acoustic Resonator (LOBAR) devices for fundamental shear horizontal modes (SH0). Detailed in Fig.~\ref{fig:wide1}(a), these devices generate acoustic waves in the center of a rectangular region, allowing several overtone modes of resonance to occur \cite{lu_lithium_2018}. Further details on stack fabrication, film quality, and device parameters can be found in Supplemental Figures 1 and 2. This gives rise to the admittance plot seen in the bottom left in Fig.~\ref{fig:wide1}(c), where each peak corresponds to a different lateral resonance of the device. Fig.~\ref{fig:wide1}(b) shows examples simulated using finite element analysis (FEA) in COMSOL Multiphysics, featuring the 9th, 17th, and 25th over modes at 86 MHz, 172 MHz, and 259 MHz. These devices are ideal candidates for parametric down-conversion, as there exists an assortment of modes in close proximity, making it much easier to design for resonances to parametrically convert from on to another. This is particularly important in our platform, where the anisotropy of LN and dispersion of Lamb waves make it challenging to fabricate devices that satisfy sum frequency generation requirements, especially at high frequencies.

\begin{figure*}
\includegraphics[width=\textwidth]{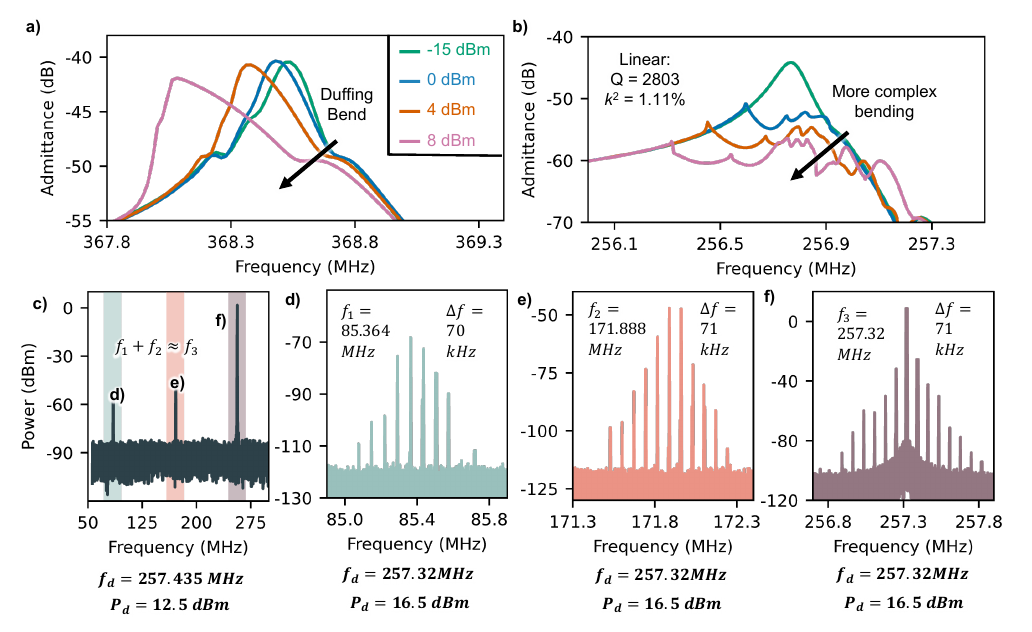}
\caption{\label{fig:wide2}(a) Chosen device driven with high power, showing only duffing nonlinearity. (b) Same device that exhibits complex behavior of resonance bending due to mode coupling for driven comb resonance. (c) Wide spectrum analysis of device driven into parametric down-conversion with $f_d$ = 256.435 MHz and $P_d$ = 12.5 dBm. (d-f) Zoom in images of each tone when driven into comb regime, showing (d) lower, (e) middle, and (f) upper tones with $f_d$ = 257.32 MHz and $P_d$ = 16.5 dBm.}
\end{figure*}

The exact mechanism of comb generation is seen in Fig.~\ref{fig:wide1}(d), showing the different regimes of device operation. The first regime, known as the linear regime, drives the device with lower power near resonance, generating weak acoustic tones only at the driven tone. When drive powers are stronger, the thermal nonlinearities start to take effect, and tones at $f_{1}$ and $f_{2}$ satisfying $f_{1}$+$f_{2}$=$f_{3}$$\approx$$f_{d}$ are generated. This process, known as parametric down-conversion, has been well-documented in micro-acoustic systems for use in amplifiers, oscillators, and comb generation \cite{ganesan_phononic_2017,raskin_novel_2000,rocheleau_micromechanical_2014}. This parametric down-conversion is a prerequisite for comb generation, as the coupling of modes is what gives rise to the nonlinear mixing that creates our comb lines. After we drive the device at even higher power, the last regime is achieved where the main and parametric tones all exhibit comb lines near their respective resonances.

The devices are additionally characterized for their magnitude of nonlinearity. In the linear regime, the driven mode is characterized by a high quality factor ($Q$) of 2803 and a moderate $k^2$ of 1.11\% stemming from a high mode number. Further details on Q and $k^2$ from modified Butterworth-van-Dyke (mBVD) model are shown in Supplemental Figure 3. LN offers high $Q$ acoustic modes, which reduces the threshold for comb generation\cite{qi_existence_2020}. When acoustic devices are driven into their nonlinear regime, their characteristic Duffing term gives rise to an amplitude-dependent frequency shift of the resonance, causing it to bend away from the nonlinear resonance, originating from joule heating\cite{segovia-fernandez_thermal_2013,lu_study_2015,landau_mechanics_1969}. As the device is driven with higher and higher power, the resonance bends further and further away, as seen in admittance measurements in Fig.~\ref{fig:wide2}(a). These sweeps are taken in reverse as to follow the direction of bending. Supplemental Figure 4 shows examples of backward and reverse sweeps to emphasize the hysteretic device behavior. It should be noticed, though, that the resonance seen for our devices contains several discontinuities, not just the one that is typically seen to occur from the Duffing term in oscillator equations. Fig.~\ref{fig:wide2}(b) shows the resonance corresponding to the tone used for driving, showing that the behavior that gives rise to comb formation leads to more mode bending than simple third order Duffing bending. It is hypothesized that these additional bifurcations occur due to out of plane motion coupling to the principal axis motion, analogous to multi-branch frequency responses reported in related nonlinear resonator systems\cite{carvalho_multiple_2017,xu_sensitivity_2024}.

Thermal modeling efforts were also conducted via COMSOL, with results showing thermal conductance values of around 2.37e-4 W/K in air and 1.46e-5 W/K in vacuum. These results are plugged into the thermal circuit model adapted from Ref.~\onlinecite{lu_study_2015}. Here, our mBVD is used to predict a temperature rise due to joule heating and to shift the resonance frequency accordingly. This allows us to plot the temperature rise along the admittance plot, showing a temperature rise on the scale of 30°C for an input power of 12 dBm (Located in Supplemental Figure 5). Efforts to measure these temperature rises are currently in progress, with reflective camera measurements being difficult in transparent films, efforts are focused more on measuring changes in meandering resistance structures. Alongside thermal simulations, results from vacuum measurements, shown in Supplemental Figure 6, show a drastic increase in nonlinearity corresponding to the drop in thermal conductance. As the device holds more heat, it becomes much more nonlinear at a given power level in vacuum than in air. Accordingly, it is shown that our comb can be achieved with a much lower drive power in vacuum setting. 

\begin{figure*}
\includegraphics{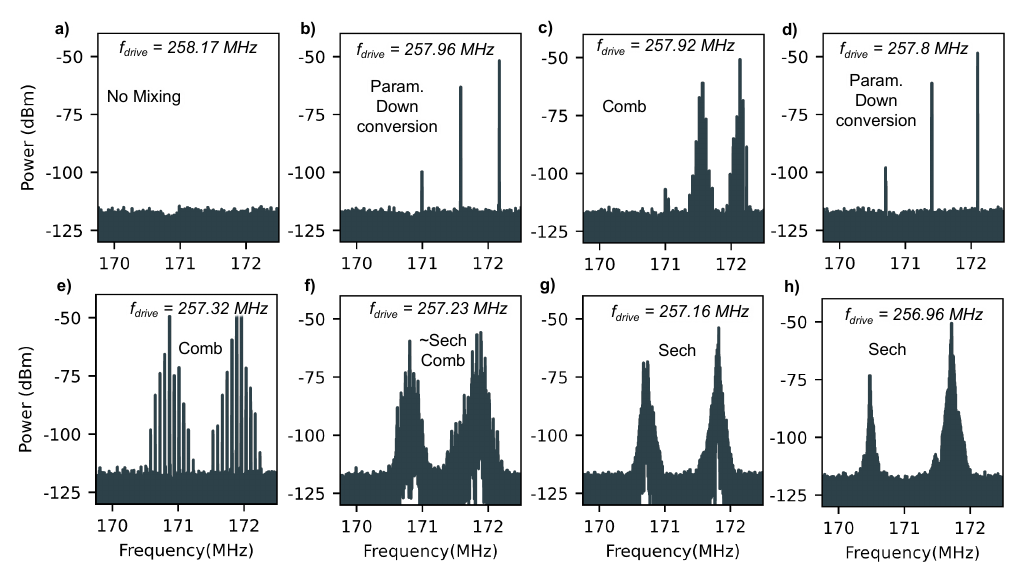}
\caption{\label{fig:wide3}(a)-(h) Different comb regimes for high power driving showing the spectrum change regimes for progressively lower frequency of drive over the resonance, all with $P_{d}$= 16.5 dBm.}
\end{figure*}

Because the devices are single-port, the reflections from the device are used to observe nonlinear phenomena. This is performed by driving the device on resonance using a signal generator and amplifier, sending the signal into the device through a unidirectional coupler, and routing the reflections back into a spectrum analyzer. For a drive frequency $f_{d}$=256.435MHz and power $P_{d}$ = 12.5 dBm, parametric down-conversion is achieved into tones at 171.9 MHz and 85.36 MHz satisfying $f_{1}$+$f_{2}$=$f_{3}$ seen in Fig.~\ref{fig:wide2}(e). While the resonator does indeed have another strong tone at 171 MHz, there is only a very weak tone at 86 MHz. Currently, the mechanism that determines exactly which tones convert energy from one to another is still under investigation, needing future theoretical models to help predict the complex mode coupling behavior.

When devices are driven at slightly lower frequency and higher power of 256.32 MHz and 16.5 dBm, combs are generated at all three tones. The frequency spacing of the comb is the same for all three tones, sitting at $\Delta f = $70 kHz. This frequency spacing is exactly equal to the difference in the sum of the first two frequencies minus that of the drive frequency. For example, here our sum is $f_1+f_2=85.364+171.888=257.252$ MHz, with a drive frequency of 257.32 MHz, this gives almost exactly $\Delta f = $70 kHz. The main tone at 257.32 MHz remains the strongest because the electrical signal is not fully attenuated by the coupler, whereas the 86 MHz tone is the weakest, likely due to a weak mechanical resonance that fails to amplify the signal. 

Because our nonlinearity arises from joule heating, small perturbations can cause large changes in device behavior, and the comb behavior in general can change greatly over the frequency of one resonance. In Fig.~\ref{fig:wide3}, the behavior is documented by changing just the drive frequency, and showing how the middle comb at 171 MHz changes in its shape. For large drive frequencies, we are too far off resonance, and no parametric mode is created. After lowering the drive frequency, we enter the parametric excitation regime, similar to phononic comb devices in literature. Also similar is the next regime, where we begin mixing into our comb. The device then begins to act chaotically, reverting to a parametric conversion regime, which is not typical. We then enter the comb regime again, and then into what is often called a "sech" or chaos comb regime, where the lines of the comb start to meld together, following the formalism in other, similar phononic comb papers\cite{zhang_optomechanical_2021,zheng_tc_2024}. In the next phase, the lines can no longer be distinguished, and the subsequent phases are merely variations on this sech/chaos function. For brevity, we stop at $f_{d}$ = 256.96, although this behavior of changing states roughly every 100 kHz persists until $f_{d}$ = 256.4 MHz, where it disappears. For lower powers, we have less Duffing bending, and the range over which we have a comb is much less; thus, a higher power was chosen to show the chaotic behavior. The regime labels in Fig. 3 are used as descriptive guides to qualitatively distinct spectral features, rather than as sharp phase boundaries defined by quantitative parameters.

\begin{figure*}
\includegraphics{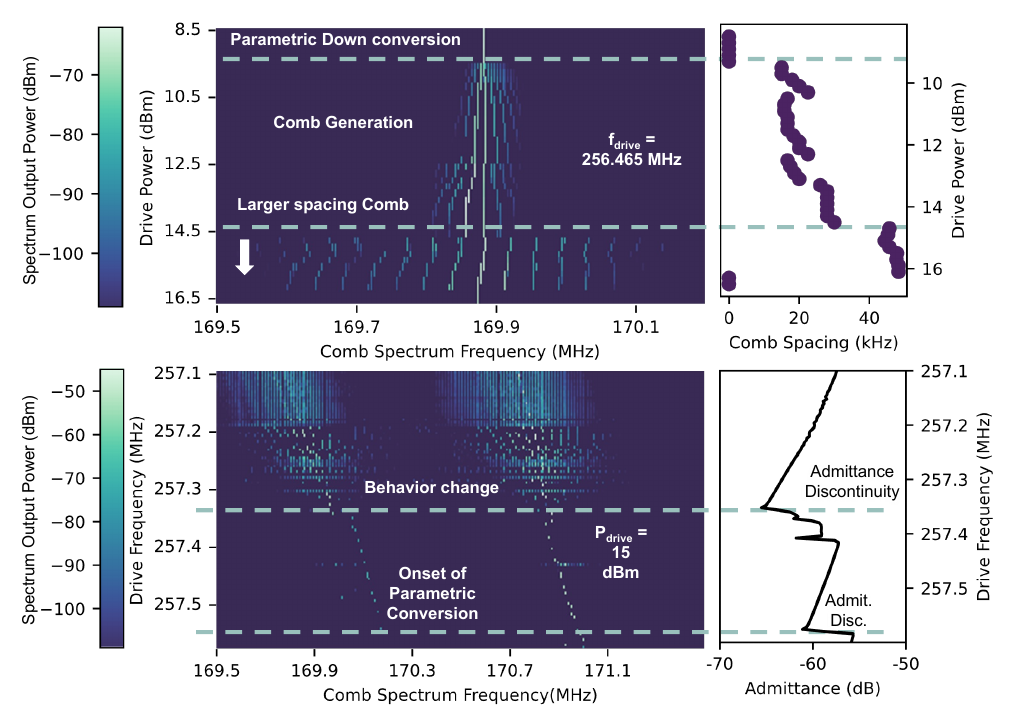}
\caption{\label{fig:wide4}(top) Heatmap featuring the spectrum at different drive powers, paired with the comb spacing at each of those powers, highlighting different regimes based on different spacings. (bottom) Heatmap featuring spectrum against drive frequency, highlighting how the behavior change will typically depend on changes in the admittance plot.}
\end{figure*}

Another piece of evidence that further reinforces the complex behavior is the presence of the second comb right next to the first (though there still exists only one set at the main tone). This second set of combs also satisfies the condition of comb spacing being equal to the difference in drive versus the sum of the two frequencies. To account for this, when considering the combs at 86 MHz and 171 MHz, the first, slightly lower frequency comb at 86 MHz is likely generated with the second, slightly higher frequency comb at 171 MHz (and vice versa). This behavior is likely similar to the phenomenon in Ref.~\onlinecite{ganesan_towards_2017}, where depending on the drive frequency, conversion into different sets of parametric tones is possible. This occurs when the frequency shift is so large that a different set of tones is also able to simultaneously satisfy our$f_{1}$+$f_{2}$=$f_{3}$ condition. However, it appears that with our devices, both combs always exist simultaneously. This second comb has $f_1$ = 86.45 MHz, $f_2$ = 170.868 MHz, giving a comb spacing of 2 kHz; thus, the comb arises from the first set of tones rather than this one. For other comb regimes, it is the second set of parametric tones that gives rise to the comb, and in some cases, both sets occur simultaneously. The exact dynamics underlying the double parametric down-conversion and comb generation behavior are intricate and not yet fully understood, and are therefore left for future investigation.

For controlled measurements, where the device is driven in a repeatable manner, as with computer-aided driving and measuring (i.e., MATLAB code for instrument control), measurements are consistent. Though it should also be noted that the device behavior is quite sensitive to environmental or setup changes. Fig.~\ref{fig:wide4} shows a heat map of the device spectrum at the middle tone for many drive powers. For low drive powers, the device operates in the parametric regime, and as power increases, it transitions into the comb regime. Within the first comb regime, it can be observed that the spacing jumps periodically but is generally increasing, a characteristic of other combs in literature. At even larger powers, the spacing suddenly jumps to a much larger value. It is believed that this occurs due to complex mode coupling, similar to the behavior in Ref.~\onlinecite{ganesan_excitation_2018} where two beams are coupled and produce a similar comb of first smaller then larger spacing. Unlike the previously mentioned double parametric down-conversion, this process is more closely related to the presence of two modes near the drive frequency. Other discontinuities in the comb spacing are likely due to discontinuities in the admittance plot or some smaller coupling dynamics, but notably, the trend is similar to that of other devices. In terms of device repeatability, various devices showed very similar comb phenomenon. Supplemental Figure 7 highlights one such case, showing a similar heatmap to previous devices, where the only difference with this device is a slight offset in electrode placement.

In the bottom of Fig.~\ref{fig:wide4} is a similar heatmap, this time plotting the spectrum against drive frequency instead of drive power. This shows a different regime compared to the first heatmap, where the behavior of the comb is significantly more sporadic, exhibiting characteristics such as the sech comb. Alongside is the admittance of the device at this drive power, showing how changes in the comb behavior will tend to correspond with abrupt changes in admittance. Investigation of the spectral broadness is quantitatively carried out via plotting the number of comb lines versus the input power and frequency shown in Supplemental Figure 8, where an optimal frequency of 256.5 MHz and power of 15.3 dBm gives the largest number of 28 comb lines.

In conclusion, we have demonstrated the first phononic combs in lithium niobate, utilizing thermal nonlinearities. LOBAR devices were used for their high density of closely spaced modes, which gives a larger probability of mode conversion. These devices were tested and found to have additional discontinuities in admittance, in addition to Duffing bending. When driven with high power, it was found that certain tones gave rise not only to parametric down-conversion but also to an array of comb types, dependent on the drive power and frequency. Future work should focus on several metrics. LDV testing of our lamb wave devices will help determine which modes we are coupling energy into, enabling direct mapping of the generated acoustic modes. Modeling efforts will help identify how to drive the comb more effectively, including deriving existence conditions, comb bandwidth, and line spacing to harness and control the nonlinearities in the devices. Future investigations should also focus on the dual parametric down conversion process, including the physical underlying mechanism and assessing how it influences comb stability and spectral response of the device. Phononic combs, like their photonic counterpart, harness their wide frequency characteristics to create short-time domain signals. While exact mechanisms are still unknown, we believe these broad frequency combs will find use in micro acoustic devices through applications such as energy harvesting or range finding.

\section*{SUPPLEMENTARY MATERIAL}
See the supplementary material for fabrication methods, details on RF characterization of devices, details on thermal and nonlinear modeling methods, comparison of atmosphere and vacuum measurements, example measurements of similar devices, and spectral broadness analysis.

\begin{acknowledgments}
This work was supported in part by the NASA Space Technology Graduate Research Opportunity (NSTGRO) grant No. 80NSSC24K1369, and the Defense Advanced Research Projects Agency Higher-Order Composite Resonators for Extra Resilience (HORCREX) program No. \ HR0011-25-9-0159. The authors appreciate Dr. Sunil Bhave, Dr. Andrea Al\`u, Dr. David Burghoff, Dr. Neal Hall, Dr. Seungwhi Kim, and Dr. Meng Kang for helpful discussions. Any opinions, findings, conclusions, or recommendations expressed in this material are those of the author(s) and do not necessarily reflect the views of the Defense Advanced Research Projects Agency (DARPA).
\end{acknowledgments}
\section*{Author Declarations}
\subsection*{Conflicts of Interest}
The authors have no conflicts to disclose.
\subsection*{Author Contributions}
\textbf{Ian Anderson:} Data Curation (lead); Conceptualization (equal); Writing – original draft (lead). \textbf{Jack Kramer:} Visualization (Equal); Writing – review \& editing (equal). \textbf{Tzu-Hsuan Hsu:} writing – review and editing (equal); Conceptualization (equal). \textbf{Yinan Wang:} Conceptualization (equal). \textbf{Vakhtang Chulukhadze:} Writing – review \& editing (equal). \textbf{Ruochen Lu:} Supervision (lead); Writing – review \& editing (equal).

\section*{Data Availability Statement}
The data that support the findings of this study are available from
the corresponding authors upon reasonable request.

\section*{}

\bibliography{comb_paper}

\end{document}